\documentstyle[pra,multicol,aps,graphicx]{revtex}

\begin{document}
\draft

 \title{On the Inverse Problem for the Multiple Small-Angle Neutron
 Scattering }

\author{D.N. Aristov $^{1,2}$ \footnote{e-mail : aristov@nordita.dk} }
\address{ $^1$ Petersburg Nuclear Physics Institute,
   Gatchina, St. Petersburg 188300, Russia \\
$^2$ NORDITA, Blegdamsvej 17, DK-2100, Copenhagen, Denmark}
 \date{} 
\maketitle

 \begin{abstract}
We consider the small-angle multiple neutron scattering and a possibility of
its model-free analysis by the inverse problem method. We show that the
ill-defined inverse problem is essentially regularized by the use of a planar
detector of neutrons without a ``beam-stop'' facility. We analyze the obtained
expressions with the examples of different models of scatterers.
  \end{abstract}

\pacs{ Keywords ~:  multiple small-angle neutron scattering, inverse problem }

\begin{multicols}{2} \narrowtext

The small angle neutron scattering (SANS) technique is an important tool for
studying large-scale inhomogeneities in condensed matter. The profile of the
intensity of passing neutrons around the $(0,0,0)$ reflection provides an
information about the size and shape of the inhomogeneities. As discussed,
e.g., in \cite{SaBer}, the procedure of extracting this useful information may
be complicated by different factors, the depletion of the incident beam and
multiple scattering among them. 

Different ways of analysis of the multiple scattering were proposed up to now
\cite{BHR,MJS,SaBer}, however all of them depend to some extent on the
particular assumptions. In fact, the assumptions are almost inevitable, since
the analysis of multiple SANS data deals essentially with an inverse problem,
which is mathematically ill-defined. 

In this paper we show that the degree of uncertainty in the
analysis of multiple scattering can be substantially reduced, when one uses
the neutron detector without a so-called ``beam-stop'' facility. This latter
item is implemented in many experimental setups in order to decrease the
presumably useless contribution of neutrons passed without a collision. We
argue that when a notable part of neutrons scatter
 in the specimen  a few times, this usually discarded intensity can provide a
crucial ingredient for the model-free data analysis. Mathematically, taking
into account the intensity at the zero angle corresponds to the substantial
regularization of the inverse problem found in the multiple SANS. 

The rest of this paper is organized as follows.  First we rederive the basic
equation known in the Bethe-Moli\`ere theory of multiple scattering. Our
treatment here is similar to one in \cite{SaBer}.  We show an intermediate
formula which is more suitable for the desired inversion.  Formally inverting
this formula, we discuss the obtained expression to some detail.  
The form of functions  expected  in some models of scattering
is shown and discussed in the last part of the paper. 

The scattering probability is determined by the single scattering cross section
$\widetilde \sigma = d\sigma/d\Omega$ with thee momentum transfer $\bf Q$. 
Given the momentum of the incident neutron $k_i$, it is convenient to
define the transverse ``momentum'' ${\bf q}= {\bf Q }/k_i$ and the normalized
quantity $ \sigma({\bf q}) = \widetilde
\sigma({\bf q}) / (\int d^2{\bf q} \widetilde \sigma({\bf q})) $.
The value $\sigma({\bf q})$ is a conditional probability of the scattering at
the angle $\bf q$. The total scattering cross section $\sigma_{0} = \int
d^2{\bf q} \widetilde \sigma({\bf q})$ enters the definition of a mean-free
path, $l$, in a usual way, $l = (\sigma_0 n_i)^{-1}$, where $n_i$ is the
scatterers' density. 

In the  diffractional limit, when $l$ is much
larger than the typical  size of inhomogeneities,  the probability of
multiple scattering is given by a Poisson distribution so that  $n$ scattering
events on the path $L$ are expected for the fraction $p_n$ of incoming
neutrons,  $p_n = (L/l)^n \exp(-L/l)$.  
The part $p_0 = \exp(-L/l)$ corresponds to the non-scattered
neutrons.

The total multiple scattering intensity is given by a sum
     \begin{eqnarray}
     I({\bf q},L) &=& \sum_{n=0}^\infty p_n I_n({\bf q})
     \end{eqnarray}
with the recursively defined $I_n({\bf q})$ : 
     \begin{eqnarray}
     I_0({\bf q}) &=& \delta ({\bf q})
     \\
     I_n({\bf q}) &=& \int d^2{\bf k}\,  I_{n-1}({\bf q}-{\bf k})
     \sigma({\bf k}) , \quad n\geq 1
     \end{eqnarray}

\noindent
In terms of 2D Fourier transform $A({\bf r}) = \int d{\bf q}
e^{i{\bf qr}} A({\bf q})$ we rewrite
     \begin{eqnarray}
     I({\bf r},L) &=& \sum_{n=0}^\infty p_n I_n({\bf r})
     \label{4}
     \\
     I_n({\bf r}) &=& [\sigma({\bf r})]^n, \quad n\geq 0
     \end{eqnarray}

\noindent
The sum (\ref{4}) is explicitly represented in the form
     \begin{eqnarray}
     I({\bf r},L)
     &=& \exp(-L/l) \sum_{n=0}^\infty
     \left(\frac Ll \sigma({\bf r}) \right)^n
     \\
     &=& \exp
     \left[
     -\frac Ll\left(1- \sigma({\bf r}) \right) \right]
     \label{theo-inten}
     \end{eqnarray}

Note that, according to our definition, $\sigma({\bf r}=0)=1$. 
Let us assume now that the scattering is isotropic, i.e., $\sigma({\bf q})$
does not depend on the direction of $\bf q$.   It follows then that  the
quantity $\sigma({\bf r})$ depends only on the absolute value of ${\bf r}$,
and we can transform (\ref{theo-inten}) by going to the ``momentum''
representation and integrating over the angle in ${\bf r}-$plane. As a result,
we obtain

     \begin{eqnarray}
     I(q,L)
     &=&
     \int_0^\infty \frac{r\,dr}{2\pi}
     J_0(qr)
     \exp \left[
     -\frac Ll\left(1- \sigma(r) \right) \right]
     \label{Be-Mo}     \\
     &=&
     \delta({\bf q}) e^{-L/l} +
     \int_0^\infty \frac{r\,dr}{2\pi}
     J_0(qr)
     \left[ e^{\sigma(r) L/l}
     - 1
     \right] e^{-L/l}
     \label{Be-Mo2}
     \end{eqnarray}
with the Bessel function $J_0(x)$. In the second line of (\ref{Be-Mo})
the integral is regularized by explicitly subtracting the singular
contribution of the direct beam.  One easily verify the equivalence of
the expression (\ref{Be-Mo}) to the Bethe's formula for the
small-angle scattering. \cite{Bethe}

The Eq. (\ref{Be-Mo}) has two disadvantages. First, it demands
the isotropy of the scattering in the $xy-$plane, which could be
violated, e.g.,  for the textured compounds.
Second, Eq. (\ref{Be-Mo}) is rather
awkward, which makes hard its subsequent analysis with taking into the
account the experimental resolution function. This latter shortcoming of
(\ref{Be-Mo}) cannot be cured if one works experimentally with the
stripe (one-dimensional) neutron dectector which was a common situation
a few decades ago.  On the other hand, the most of the neutron centers
are equipped nowadays by two-dimensional (2D) SANS machines, and it
provides a possibility for the easier data analysis as explained below.

Keeping in mind the 2D experimental geometry, we use the Eq.
(\ref{theo-inten}), rather than its particular form (\ref{Be-Mo}).
Next we consider the experimental situation when the idealized
intensity $I({\bf q},L)$ is convoluted with the instrumental resolution
function $F({\bf k})$. For simplicity, we assume that this is the Fourier
convolution. In addition, we take into account that some part
of neutrons could be lost due to absorption or incoherent scattering.
These effects are irrelevant to the discussed small-angle scattering
and may be modelled by the unique absorption length, $l_a$.  The
measured (experimental) intensity $I_{exp}({\bf q},L)$ is then given by
     \begin{eqnarray}
     I_{exp}({\bf q},L) &=& e^{-L/l_a}
     \int d^2{\bf k}\,  I({\bf q}-{\bf k},L)
     F({\bf k}) ,
     \end{eqnarray}
or
     \begin{eqnarray}
     I_{exp}({\bf r},L) &=& e^{-L/l_a}  I({\bf r},L) F({\bf r}).
     \label{exp-inten}
     \end{eqnarray}

\noindent
At $L=0$ one has $I({\bf r}) =1 $ and the obvious equality
$I_{exp}({\bf r},L=0) = F({\bf r})$, i.e. without a sample one measures
only the shape of instrumental resolution.  

It should be stressed here, that the next step of our analysis will require a
certain conservation rule. Specifically, we observe that in the absence of
absorpition ($L/l_a \ll 0$), one has $I_{exp}({\bf r}=0,L) = 1$, which
equality means simply that the total number of neutrons, scattered and passed
without collision, does not depend on the thickness $L$. This
obvious conservation law is violated, when one excludes from the
consideration the forward-going neutrons. In practice, it is done
by the conventional setup with a ``beam-stop'', which assumes particularly that
the forward-going neutrons carry no useful information. We argue here that the 
part $I_{exp}({\bf q} \simeq 0,L) $ is very important, enabling one 
to analyze the multiple SANS without further adjustable parameters.

Indeed, given our assumptions above, 
we see that if the quantity $I_{exp}({\bf r},L)$ is determined for 
two samples with thicknesses $L_1$, $L_2$, then 
the ratio $I_{exp}({\bf r},L_1) /I_{exp}({\bf r},L_2)$ does not  depend on
$F({\bf r})$. 
It allows us to invert the
equations (\ref{theo-inten}), (\ref{exp-inten}) and to write
     \begin{equation}
     \frac1{L_2 -L_1} \log\frac {I_{exp}({\bf r},L_1)}
     {I_{exp}({\bf r},L_2)} =
     \left[ \frac {1- \sigma({\bf r}) }l + \frac1{l_a} \right]
     \equiv f({\bf r})
     \label{main}
     \end{equation}

\noindent
This equation is the main result of our paper.
Let us discuss it in more detail.

1) The rhs of Eq.(\ref{main}) does not depend on the sample thickness
$L_{1,2}$, therefore one can combine the scattering data, obtained for
different $L_i$ and to improve the determination of the shape of
$f({\bf r})$.

2) For the case of the isotropic scattering,
$\sigma({\bf r}) = \sigma(r)$, one averages over the directions in
$xy-$plane in (\ref{main}) and not in (\ref{Be-Mo}).

3) Generally, $I_{exp}({\bf r},L)$ is a complex-valued function.
However, the ratio $I_{exp}({\bf r},L)/I_{exp}({\bf r},0)$ entering
(\ref{main}) should be real, provided $\sigma({\bf k}) = \sigma(-{\bf k})$.
This condition means simply the inversion symmetry which is generally present
on the SANS scale (violating probably in some special
cases, e.g. for some substances in the magnetic field).  
Adopting  that $\sigma({\bf k}) = \sigma(-{\bf k})$, we note that the
imaginary part of logarithm in (\ref{main}) can stem from two reasons. The
first reason is the statistical noise and the second one is the displacement
of the direct beam (${\bf q}=0$) during the measurements with different $L_i$.
If ${\bf q}_{disp} $ is the displacement of the direct beam, then $I_{exp}({\bf
r},L) / I_{exp}({\bf r},0)$ acquires the factor $\exp(i{\bf q}_{disp}
{\bf r})$.  This factor can be easily determined by the appropriate
analysis of the data. The remaining complexity in the value of the
discussed ratio is due to the statistical noise. Both discussed sources
of the complex quantity in the lhs of (\ref{main}) are irrelevant to
the physical quantity $\sigma({\bf r})$ and can be stripped off by
writing
     \begin{equation}
     f_{reg}({\bf r}) =
      \frac1{L_2-L_1} \log\left|\frac {I_{exp}({\bf r},L_1)}
     {I_{exp}({\bf r},L_2)}\right|
     \label{main-reg}
     \end{equation}

\noindent
This definition of $f_{reg}({\bf r}) $  is the
partial {\em regularization} of our inverse problem, since it projects
the experimental data into a definite (symmetrical) class of the
functions $\sigma({\bf q})$.  Hereafter we assume that $f({\bf r})$ is
regularized according to (\ref{main-reg}) and drop the subscript in the
function $f_{reg}({\bf r})$.
In fact, the  regularization (\ref{main-reg}) is not very restrictive. 
The crucial regularization of the problem appeared in
(\ref{exp-inten}),  corresponding to the assumption of 
the planar detector without "beam-stop". 

4) Two basic features in the determined shape of $f(r)$
are of interest.  First, since $\sigma(0)=1$ one has  $f(0) = 1/l_a$, i.e.\ 
the absorption length is found in the same experiment as the
other quantities.  Second, one expects that $\sigma(\infty) =0$ (see
below) so that $f(\infty) =1/l+ 1/l_a$ and hence the value of $l$ is
determined.

5) Knowing $l$ and $l_a$, we find $\sigma(r)$ and can compare the form
of this function with the different models 
of  scatterers, partly discussed below. The characteristic scale $r_0$ of the
variation of $\sigma(r)$ is connected with the size of the inhomogeneities $R$
by a simple relation $R = r_0/k_i$.

6) The above limit
$\sigma(\infty)=0$
corresponds to the
vanishing angle between the neighboring neutron detectors, which is
evidently never realized. In practice, one deals with the finite Fourier
transform, and the finite-size effects appear at largest $\bf r$,
determined by the resolution $ \Delta q$.  One may judge confidently about the
limiting value $f(\infty)$ when the curve $f(r)$ saturates at
$r<1/\Delta q$. 
This saturation takes place when the average size of the scatterers is smaller 
than $1/\Delta q$. 
In this case one uniquely determines two main parameters $l$
and $r_0$, along with the whole ${\bf r}-$dependence of $\sigma({\bf r})$.

7) It was proposed in \cite{MJS} to analyze the multiple scattering by
inverting Eq.(\ref{Be-Mo2} with the subtracted $\delta-$function. We note here
that such procedure requires the knowledge of the ratio $L/l$ and should be
sensitive to its choice. The inversion of (\ref{Be-Mo2}) is also vulnerable
due to the ignorance of the absorption, $l_a$, and the experimental
resolution. 

It is worth to provide here  some examples of the possible shapes of
$\sigma(r)$. We do it first  for  
the scattering on the spheres and for the phenomenological
``Ornstein-Zernike-squared'' cross-section. 

In the first case, one has the spherical scatterers of one radius $r_0$, and 
 the differential cross-section in the diffraction limit reads as
     \begin{equation}
     \sigma_{sph}(q) \propto  q^{-3} J_{3/2}^2(qr_0)
     \label{sph-q}
     \end{equation}
For this function one finds the image $\sigma(r)$ in the form
     \begin{equation}
     \sigma_{sph}(r) = g_{sph}(r^2/4 r_0^2)
     \label{sph-r}
     \end{equation}
with \cite{BHR}
        \begin{eqnarray}
        g_{sph}(x) &=&
        \sqrt{1- x}\left(1+\frac{x}2 \right)
        \nonumber \\ &&
        - x \left(2-\frac{x}2 \right)
        \ln \left[ \frac{1+\sqrt{1- x}}{\sqrt{x}} \right]
        \label{spher}
        \end{eqnarray}
at $x \leq 1$ and $g_{sph}(x)=0$ otherwise. 

In the second case, one writes 
     \begin{equation}
     \sigma_{OS2}(q) \propto (q^2 r_0^2 + 9/2 )^{-2}
     \label{OS-q}
     \end{equation}
with the choice of the factor $9/2$ explained below. The image $\sigma(r)$ is 
found as
     \begin{eqnarray}
     \sigma_{OS2}(r) & =&  g_{OS2}(r/2r_0) 
     \label{OS-r} \\
      g_{OS2}(x) &=& (3 x/\sqrt{2}) K_1(3x/\sqrt{2}),
     \end{eqnarray}
with the modified Bessel function $K_1(x)$.  We note that the expression
(\ref{OS-q}) corresponds to the Guinier radius $ R_g^2 = 4 r_0^2/3$, which is 
approximately two times larger than in the case of spheres, $R_g^2 =
3r_0^2/5$. 

\begin{figure}[ht]
\includegraphics[width=7cm] {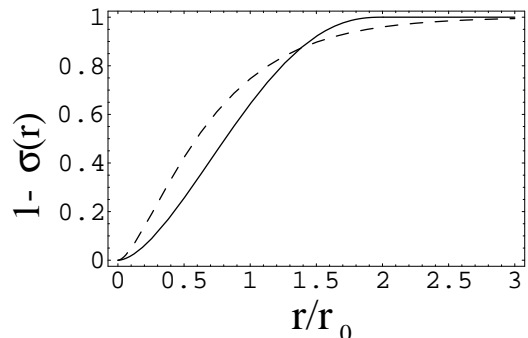}%
\caption{Two examples of possible functions in eq. (\ref{main}). The profiles 
described by eq. (\ref{sph-r})  and eq. (\ref{OS-r}) are shown by the solid
and dashed lines, respectively. } 
\label{fig:1} 
\end{figure}

The behavior of the corresponding functions $(1-\sigma(r))$, entering eq.\ 
(\ref{main}), is shown in Fig.\  \ref{fig:1}. An obvious saturation at
large $r$  takes place in the case of spherical scatterers, while this feature
is not so prominent  for the function $(1-\sigma_{OS2}(r))$. On the other
hand, the point of intersection of both curves at $(1-\sigma(r)) \simeq 0.88$
at $r\simeq 1.4 r_0$ turns out to be  semi-invariant. 

To explain the last statement, we remind that 
in practice one can
rarely expect that all the scatterers are of the same size. To model
the situation, we suggest that the sizes of the spherical scatterers
are distributed within some range, with the probability $P(R)$ to find a
sphere of radius $R$. The function $P(R)$ is normalized and the mean
square of $R$ is $r_0^2$  :
        \begin{equation}
        \int_0^\infty P(R) dR =1
        ,\qquad
        \int_0^\infty R^2 P(R) dR = r_0^2
        \label{normaliz1}
        \end{equation}
Then one can write
     \begin{equation}
     \sigma(r) =
     \int_0^\infty dR\, g_{sph}(r^2/4R^2) P(R)
     \label{sph-distr1}
     \end{equation}
The simplest one-parameter family of functions $P(R)$, obeying
(\ref{normaliz1}), is given by 
    \begin{equation}
     P_\alpha(R) = \frac{2}{R} F_\alpha \left(\frac{R^2}{r_0^2} \right),
     \qquad
     F_\alpha(x) =
     \frac{(x\alpha)^\alpha}{\Gamma[\alpha]}
     e^{-x\alpha}
     \label{probab2}
     \end{equation}
The parameter $\alpha$ characterizes the width of
distribution, according to a relation $\langle R^4 \rangle - \langle R^2
\rangle^2 = r_0^4/\alpha$, with $\langle \ldots \rangle$ standing for the
average. 
Thus the case  $\alpha\to \infty$ corresponds to the previous case of the
monodispersed situation, $P_\infty(x) = \delta(R-r_0)$, while 
$P_\alpha(x)$ becomes uniform distribution in 
the limit $\alpha\to 0$.  

Substituting (\ref{probab2}) into (\ref{sph-distr1}),  we find the family 
of curves $\sigma_\alpha(r)$ in the form
      \begin{equation}
      \sigma_{\alpha}(r) = \frac{\sqrt{\pi}}{\Gamma[\alpha]}
      G^{30}_{23}\left(\left. \frac{r^2 \alpha}{4r_0^2}  \right|  
       \begin{array}{l}         3/2,3 \\        0,1,\alpha        \end{array} 
      \right),
      \label{sigma-fam} 
      \end{equation} 
with the Meijer function, implemented, e.g., in {\em Mathematica}. 
We numerically found that the above point 
$(1-\sigma(r)) \simeq 0.88$ and $r\simeq 1.4 r_0$ 
is semi-invariant for the family (\ref{sigma-fam}) 
 for $0.4\leq \alpha \leq 15$.  For completeness, we also present here the
form of $\sigma(q)$. Up to the overall coefficient one has  : 

      \begin{equation}
      \sigma_{\alpha}(q) = q^{-2}
      G^{21}_{32}\left(\left. \frac{ \alpha}{q^2r_0^2}  \right|  
       \begin{array}{l}   0, 3/2,3 \\   1,\alpha  \end{array} \right),
      \label{sigmaq-fam} 
      \end{equation} 

Let us now discuss our choice of the factor 9/2 in Eq. (\ref{OS-q}). 
This factor appears quite naturally when one poses oneself a
following question. Which distribution $P(s)$  in (\ref{sph-distr1}) leads to
the intensity profile of the form $(q^2  r^2 +1)^{-2}$ and what is the
relation between $r_0$ and $r$ here ?

\begin{figure}[ht]
\includegraphics[width=7cm] {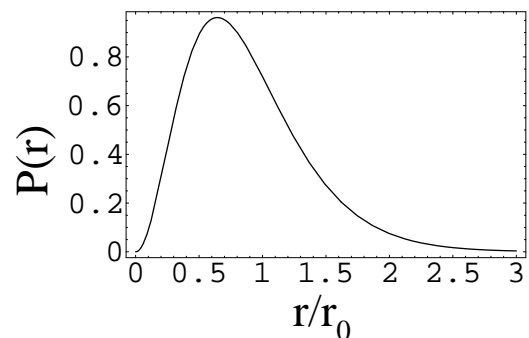}%
\caption{The distribution function of the spherical scatterers, leading to the 
 intensity (\ref{OS-q}). The function $P(r)$ is measured  in units 
of $r_0^{-1}$.  }  \label{fig:2} 
\end{figure}

The answer to this question is based on the observation, that
eq. (\ref{sph-distr1})  is the Mellin convolution of the functions $g_{sph}(x)$
and $P(x)$ and, in principle, may be inverted for the known $\sigma(r)$.  
The corresponding derivation  involves the theory of Mellin-Barnes
integrals, discussed, e.g. in \cite{Marichev}. Without going into details, we
quote here the final result of the form 
            \begin{eqnarray}
            P(R) &=& (2/R) F(R^2/r_0^2) \nonumber \\
            F(x) &=& \left( \frac{9x}2 \right)^2 e^{-3\sqrt{2x}} 
             \left( 1+ \frac1{3\sqrt{2x}}\right).
            \label{P-tran}
            \end{eqnarray} 
The behavior of this function is shown in Fig.  \ref{fig:2}. Therefore 
 the distribution (\ref{P-tran}) of spherical scatterers with average 
$\langle R^2 \rangle =r_0^2$ leads to the  intensity profile (\ref{OS-q}),
(\ref{OS-r}). 

Consider also a situation when the inhomogeneities in the specimen form
clusters, whose size is described  by fractal statistics. \cite{Ma-fra} In
the case of simple volume fractals, the commonly adopted density-density
correlation function is given by                      
          \begin{equation}
         \langle \rho(r) \rho(0) \rangle \propto
          r^{D-3} \exp(-r/\xi),
          \label{dd-fra}
          \end{equation}
with a fractal dimension $D$ and maximum cluster size  $\xi$. Given the
function (\ref{dd-fra}), one finds the form of the single scattering cross
section $d\sigma/d\Omega$, exhibited elsewhere  \cite{Ma-fra}. Its 
counterpart in $r-$space, entering our eq. (\ref{main}) reads  as 
        \begin{equation}
      \sigma(r) = 
      G^{30}_{13}\left(\left. \frac{ r^2}{4\xi^2}  \right|  
       \begin{array}{l}   1/2 \\   0, \frac{D-2}2 , \frac{D-1}2 
       \end{array} \right),       
       \label{sigmar-fra} 
      \end{equation} 
The saturation of this function at large $r$ is realized when $r>\xi$.  In
particular, it means that one cannot uniquely determine the parameters
$\xi$, $l$, $l_a$ in the case when $\xi \Delta q \agt 1$.

In conclusion, we discuss the inverse problem for the multiple small-angle
neutron scattering. We show that the consideration of  the forward-going
neutrons corresponds to the essential regularization of this problem. The
proposed  formalism may be applied in the region where 
the sample thickness is of order of  a few mean-free paths $l$.

\acknowledgements

I thank S.V. Maleyev, B.P. Toperverg, S.V. Grigoriev, J.Teixeira, G. Pepy for
various useful discussions. The partial financial support from the  RFBR Grant
00-02-16873, Russian State Program for Statistical Physics (Grant VIII-2)  and
Russian Program ``Neutron Studies of Condensed Matter'' is acknowledged. 

%
%
%


\end{multicols}

\end{document}